\documentclass[iicol]{sn-jnl}

\usepackage{epsfig,amssymb,amsfonts,amsmath,mathtools,bm,color,graphicx,orcidlink,bm,braket,multirow,dsfont,mathtools,xcolor,slashed}

\RequirePackage[numbers,sort&compress]{natbib}
\newcommand{\be}{\begin{equation}}
\newcommand{\ee}{\end{equation}}
\newcommand{\bea}{\begin{eqnarray}}
\newcommand{\eea}{\end{eqnarray}}
\newcommand{\beas}{\begin{eqnarray*}}
\newcommand{\eeas}{\end{eqnarray*}}

\graphicspath{{Figs.dir/}}

\begin{document}

\title{Why fluctuations of conserved charges in the confining regime
above $T_{ch}$ behave as if the quarks were free?}

\author{L.~Ya.~Glozman}
% \thanks is optional - remove next line if not needed
%\thanks{\emph{Present address:} Insert the address here if needed}%
                     % Do not remove
%

\affil{\orgdiv{Institute of Physics}, \orgname{University of Graz}, \orgaddress{ \postcode{8010}, \city{Graz}, \country{Austria}}}

%\PACS{{12.38.Aw}{} \and {12.39.Ki}{} \and {11.30.Rd}{}}

\date{}

\abstract{Some cumulants of the fluctuations of conserved charges
soon above the chiral crossover behave as if the quarks were
free. This was taken by many as evidence of 
deconfinement. At the same temperatures the
mesonic correlators reveal the chiral spin and $SU(4)$ symmetries,
indicating that the propagating degrees of freedom are massless
quarks connected into color singlets by the chromoelectric confining 
string. These correlators are qualitatively different from the free
quark gas. Here we clarify the reason for the difference. The conserved quark number densities do not
propagate in time but do propagate in spatial directions.   The mesonic propagators calculated in full QCD differ
radically from the free quark loop (quark gas) above $T_{ch}$. In contrast,
the quark number density spatial propagator in full QCD at $T \geq 220$ MeV is very close to the free quark loop. 
 In orther words,
the conserved charges do not see confinement, in contrast to the mesonic correlators. This  is consistent with the  well
understood quark-hadron duality   at $T=0$ in $e^+e^- \rightarrow hadrons$, where at invariant masses above $2 $ GeV the cross-section
in the confining regime is represented by the free quark loop plus small perturbative corrections.
All these features above  $T_{ch}$ but below the deconfinement temperature $T_d$
can be combined within the following microscopic picture of the stringy fluid matter. It is a medium of the overlapping strongly interacting color singlet
clusters. The quark interchanges between the clusters, required by  Pauli
principle, make the quarks quasifree, which is reflected in fluctuations of conserved charges. 
} %end of abstract
\maketitle

\section{Introduction}

The fluctuations of conserved charges is one of the main topics
for experimental and lattice studies in the QCD phase diagram
community. Initially they were suggested as an indicator of the
deconfinement at high tempreatures \cite{Asakawa:2000wh,Jeon:2000wg,Ejiri:2005wq}. For the last 20 years 
there have appeared a lot of lattice results on fluctuations at vanishing
and small baryon chemical potentials, see e.g. \cite{Bel,Borsanyi:2018grb,Bazavov:2017dus,Goswami:2026hit} and references therein. The conserved charges
do not propagate in time and consequently the strategy was choosen that
bypasses the  study of observables with the time evolution. Namely, the
conserved charges, e.g. the baryon charge, electric charge and strangeness 
can be directly related to the conserved quark
numbers of different flavors. The latters  can be extracted from the grand canonical partition function as the first derivative with respect to the quark chemical potentials. The expectation values of the conserved quark numbers of different flavors vanish
at zero chemical potentials. But their fluctuations do not vanish and can be calculated
as derivatives of the conserved quark numbers with respect to the chemical potentials taken at zero.  One of the results of the extensive lattice studies of the issue 
was that while cumulants of fluctuations are well described by the hadron gas
model below the chiral restoration temperature $T_{ch} \sim 155$ MeV, some of them radiacally 
deviate from the hadron gas predictions just above $T_{ch}$ and very soon (around 220 MeV) approach the value predicted by the free quark gas. This was taken by many as an indication of  deconfinement just above $T_{ch}$.

 This interpretation obviously contradicts  the emerged chiral spin and $SU(4)$
 symmetries in mesonic propagators \cite{R1,R2,R3,Chiu}, which point out the importance of
 the confining electric interaction at these temperatures (for complementary reviews on symmetries and their implications see Refs. \cite{G1,G2}). It also
 contradicts
 the results for pressure, energy density and entropy density, because a clear signal of the deconfinement and onset of
the quark-gluon plasma with quasi-free quarks and gluons is the Stefan-Boltzmann $\sim T^4$ behavior which  sets in at much larger temperatures $ T > 400 - 500$ MeV \cite{Bazavov:2017dsy}. The above interpretation is also in conflict with the pion spectral function that shows very
clear $\pi,\pi'$ peaks above $T_{ch}$ \cite{LP} as well as with the bottomonium spectrum
which remains the same as in vacuum at $T < 300$ MeV (the states become broader
with the temperature, however) \cite{Ding}. 

Given symmetries and degrees of freedom the QCD phase diagram at small
chemical potential was divided into three regimes (regions) connected
by smooth crossovers: the hadron gas at $T < T_{ch}$, the stringy fluid
at $ T_{ch} < T < 3 T_{ch}$ and the quark-gluon plasma at higher temperatutes
\cite{R2,GPP}. The three regimes differ by the $N_c$ scaling of energy density,
pressure and entropy density \cite{CG1,G4}: $N_c^0$ in the hadron gas, $N_c^1$
in the stringy fluid and $N_c^2$ in the quark-gluon plasma. A very smooth
crossover between the stringy fluid and the QGP is most probably centered
at the deconfinement temperature $T_d \sim 300$ MeV of the pure glue theory \cite{CG1,CG2},
which is supported by the deconfinement temperature in QCD extracted from
the center vortices percolation \cite{Mickley} as well as the Hagedorn temperature
$T_H \sim 300$ MeV within the string description \cite{Fujimoto,Mar}.

Recently we have published the paper \cite{CG2}, where we demonstrated
that the fluctuations of conserved charges above $T_{ch}$ scale as $N_c^1$,
and thus are not directly sensitive to the deconfined gluons. However, the issue why
the fluctuations of conserved charges behave in the confining regime above $T_{ch}$ as
if the quarks were free was not addressed.
In this note we explain the reason why the conserved charges do not see confinement
in the confining regime. We rely on the concept of the quark-hadron dulaty.
We do not calculate anything new, but  give the  missing 
in the literature correct interpretation of this observable.
We also present an outline of the microscopic picture of the stringy fluid
that is consistent with all available lattice data.

\section{Comparison of the QCD correlators with the free quark gas
correlators.}

On the lattice all properties of hadrons are encoded in their
temporal and spatial correlation functions.
E.g. the $\rho$-meson is excited from the vacuum by the vector-isovector quark-antiquark
bilinear  $O_\Gamma(t,x,y,z)=\bar{\psi}(t,x,y,z)\gamma_\mu \frac{\boldsymbol{\tau}}{2}\psi(t,x,y,z)$. 

Then the Euclidean correlation functions,
\begin{equation}
C_\Gamma(t,x,y,z)=\langle O_\Gamma(t,x,y,z)\,O_\Gamma(0,\mathbf{0})^\dagger\rangle\;,
\end{equation}
carry the full dynamical information of all isovector excitations with
the $\rho$-meson quantum numbers. 
The  temporal and spatial correlators in Euclidean space are defined as 

\begin{equation}
C_\Gamma^t(t)=\sum_{x,y,z}C_\Gamma(t,x,y,z)\;,
\label{eq:c_t}
\end{equation}

\begin{equation}
C_\Gamma^s(z)=\sum_{x,y,t} C_\Gamma(t,x,y,z)\;.\label{eq:c_z}\\
\end{equation}

The temporal correlators  (\ref{eq:c_t}) reflect dynamics of the QCD Hamiltonian
since $H$ translates states  in Euclidean time.
The masses of hadrons are extracted from the exponential decay
of the temporal correlators.
The spatial correlators are connected to the dynamics of the analogous operator $H_z$ translating states in 
$z$-direction. At zero temperature the temporal and spatial spatial correlators
supply equivalent information, while at $T > 0$ the information that is
encoded in both correlators is different.

When we extract the $\rho$-meson mass	from the temporal correlators we can use
any $\mu$ from the set $\mu=1,2,3$. The temporal component $\mu=4$ is not allowed
because it represents the conserved charge density that does not excite the
$\rho$-meson and does not propagate in time. With the spatial correlators
(\ref{eq:c_z}) only  $\mu=1,2,4$ can be used and the $\mu=3$ component does not
propagate in $z$-direction. At $T=0$ the temporal correlation functions
with $\mu=1,2,3$ are identical with the spatial correlation functions at
$\mu=1,2,4$, respectively.

The latter identity is violated upon increase of temperature. Still at
temperatures essentially below the chiral crossover, $ T << T_{ch}$,
all three spatial correlation functions with $\mu=1,2,4$ are equivalent.
Around $T_{ch}$ a drammatic rearrangement of the spatial correlators happens.
While the degenerate $\mu=1,2$ correlators still supply the information about the
$\rho$-meson screening masses, the conserved quark number density bilinear 
$\bar{\psi}(t,x,y,z)\gamma_4 \frac{\boldsymbol{\tau}}{2}\psi(t,x,y,z)$
decouples from the $\rho$-meson, with essentially larger slope of the correlator, and is
connected to observable fluctuations of conserved charges. This is true
not only for the $\bar{\psi}(t,x,y,z)\gamma_4 \frac{\boldsymbol{\tau}}{2}\psi(t,x,y,z)$ spatial correlator, but also for all
other conserved quark charges.

\begin{table}
%\center
\begin{tabular}{cccll}
\hline\hline
 Name        &
 Structure &
 Abbreviation    &
 \multicolumn{2}{l}{
   %Symmetries
 } \\\hline
 %%%%%%%%%%%
\textit{Pseudoscalar}        & $\gamma_5$                 & $PS$         & \multirow{2}{*}{$\left.\begin{aligned}\\ \end{aligned}\right] U(1)_A$} &\\
\textit{Scalar}              & $\mathds{1}$               & $S$          & &\\\hline
\textit{Axial-vector}        & $\gamma_k\gamma_5$         & $\mathbf{A}$ & \multirow{2}{*}{$\left.\begin{aligned}\\ \end{aligned}\right] SU(2)_A$}&\\
\textit{Vector}              & $\gamma_k$                 & $\mathbf{V}$ & & \\
\textit{Tensor}       & $\gamma_k\gamma_3$         & $\mathbf{T}$ & \multirow{2}{*}{$\left.\begin{aligned}\\ \end{aligned}\right] U(1)_A$} &\\
\textit{Axial-tensor} & $\gamma_k\gamma_3\gamma_5$ & $\mathbf{X}$ & &\\
\hline\hline
\end{tabular}
\caption{
A complete set of isovector $J=0,1$ operators in spatial correlators
and their
$U(1)_A$ and $SU(2)_L \times SU(2)_R$ transformation properties.  The
open vector index $k$ denotes the components $1,2,4$, \textit{i.e.} $x,y,t$.}
\label{t1}
\end{table}

\begin{figure}
\centering 
  \includegraphics[width=0.7\linewidth]{{{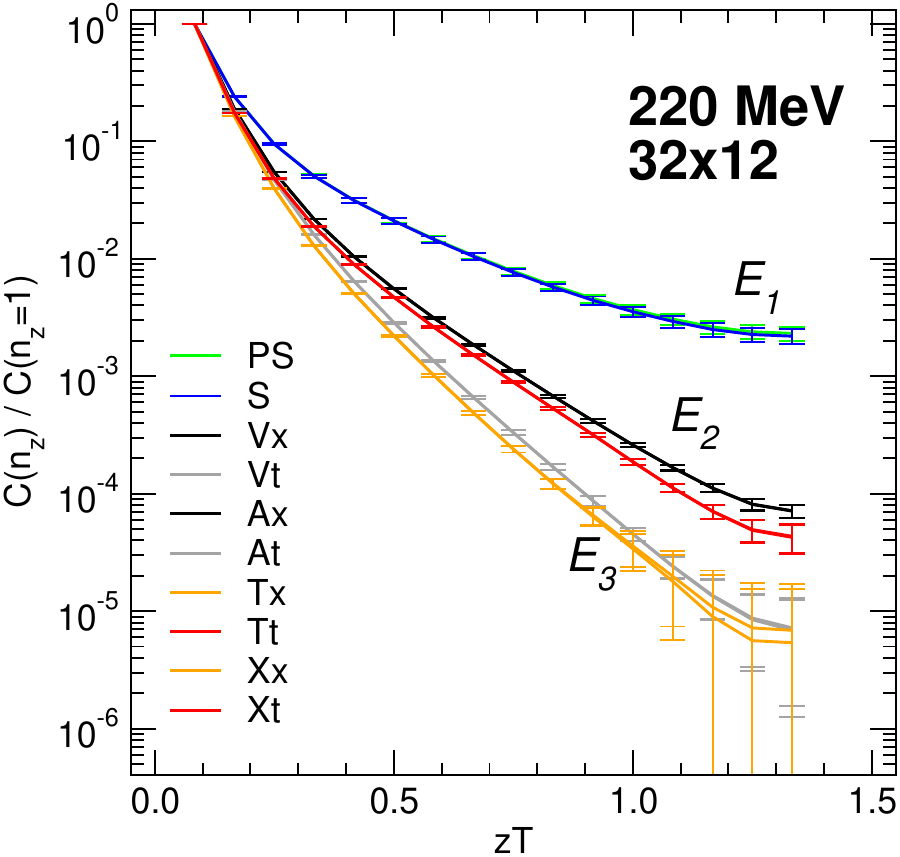}}}
  \includegraphics[width=0.7\linewidth]{{{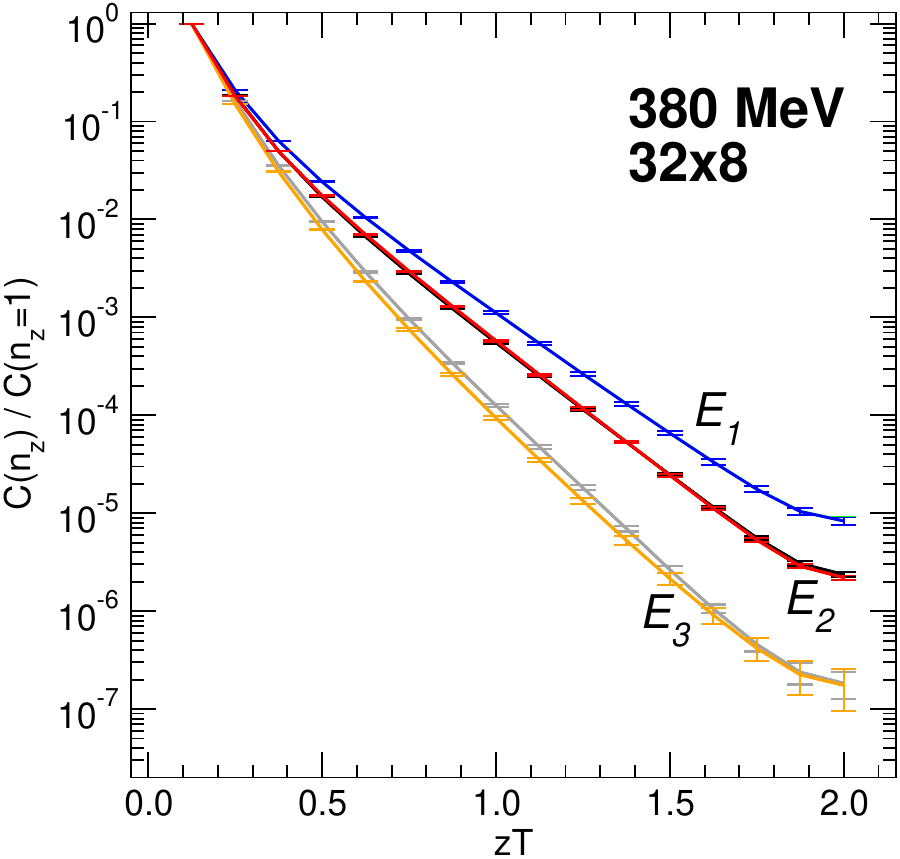}}}
\caption{Spatial
correlation functions of all possible isovector $J=0,1$ bilinears.
  From
Ref. \cite{R2}.}
\label{spatial}
\end{figure} 
To visualize this discussion we show in Fig. \ref{spatial} a complete set
of spatial isovector correlators calculated in $N_F=2$ QCD at physical quark masses
with chirally symmetric Dirac operator \cite{R2}. The set of local bilinear operators is presented in Table \ref{t1}. We see a clear multiplet strucure of the correlators at both temperatures above $T_{ch}$. The multiplet $E_1$
consists of the scalar and pseudoscalar correlators that are connected by the
$U(1)_A$ transformation. The degeneracy of both correlators indicates that the
$U(1)_A$ symmetry is approximately restored and a possible remaining $U(1)_A$ breaking due to the axial anomaly is small and cannot be seen with the presented
correlators. The multiplet $E_2$ represents  appromitale degenerate mesonic $J=1$ correlators that are connected by the $SU(2)_{CS}$ and $SU(4)$ symmetries.
The emergence of these symmetries implies that the propagating color-singlet mesons represent
the chirally symmetric quarks connected by the electric string. 
\begin{figure}
\centering 
  \includegraphics[width=0.7\linewidth]{{{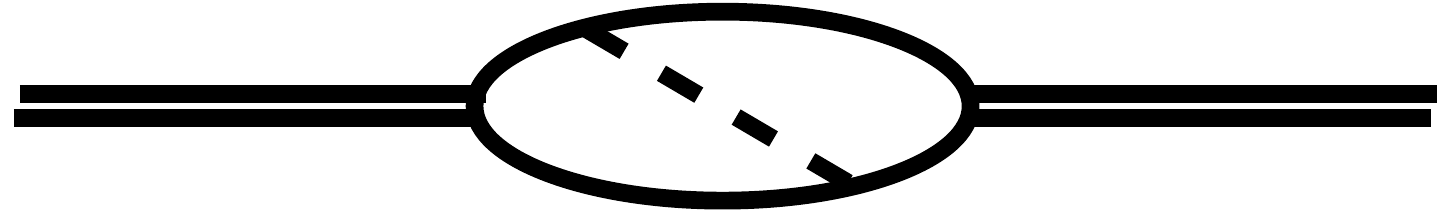}}}
  \caption{Mesonic correlator with the dashed line representing the
  confining electric string.}
\label{mesconf}
\end{figure} 
The mesonic correlators  from the $E_1$ and $E_2$ multiplets are schematically
shown in Fig. \ref{mesconf}, where the confining electric string
between the massless quark and antiquark is depicted as dashed line and the double line is the bilinear operator $O_\Gamma$ that creates the quark-antiquark pair from the vacuum at some point.

The multiplet
$E_3$ consists of four operators that  include the conserved vector and axial charge density operators $V_t$ and $A_t$. The conserved charge densities cannot propagate in
time but can propagate in z-direction.

\begin{figure}
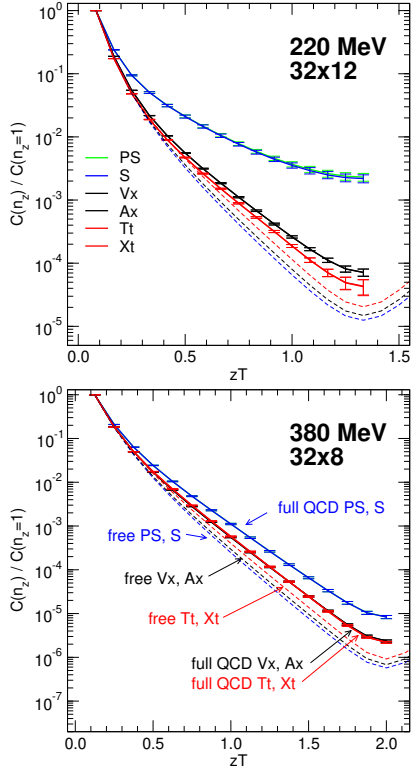

  \centering
  \includegraphics[width=0.7\linewidth]{{{2a_nt12}}} 
  \includegraphics[width=0.7\linewidth]{{{2b}}} 
 \caption{Comparison of the full QCD mesonic correlators
  from the multiplets $E_1$ and $E_2$ with with the correlators obtained
  on the same lattice with free noninteracting quarks. From Ref. \cite{R2}.}
\label{comparison}
\end{figure}

\begin{figure}
\centering 
  \includegraphics[width=0.7\linewidth]{{{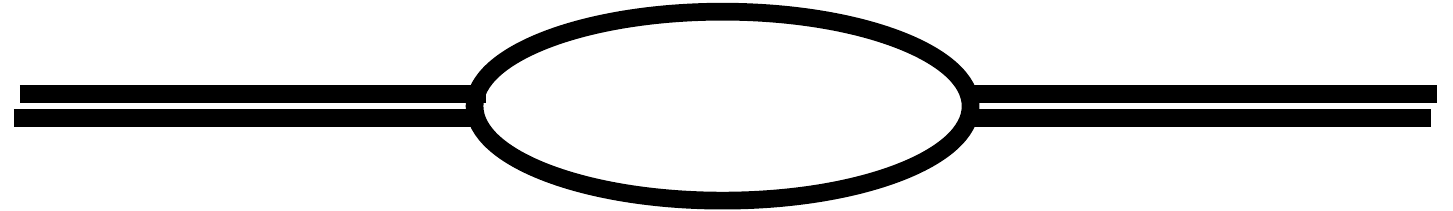}}}
  \caption{Free quark loop.}
  \label{loo}
\end{figure}

It is instructive to compare the mesonic correlators from the multiplets $E_1$
and $E_2$ in QCD with the correlators obtained on the same lattice with
noninteracting quarks, see Fig. \ref{comparison}. 
The latter correlators stand for the free
quark gas and consequently correspond to the quark-gluon plasma at a very high temperature where the quark-gluon interaction can be neglected. Diagramatically the free quark gas is given by the free quark loop, shown in Fig. \ref{loo}.
We see a dramatic  difference between the full QCD correlators and the free
quark gas, especially for the $J=0$ mesons. The pion spectral function, extracted
from the spatial correlator in Ref. \cite{LP} shows very clean $\pi, \pi'$
peaks above $T_{ch}$, which become broader with temperature. The $\rho$ bumps
with two different properties with respect to chiral symmetry, as well as the $a_1,b_1$ bumps,
that could be extracted from the corresponding correlators in the $E_2$ multiplet should be expected
to be essentially broader than the $\pi$ peak.

\begin{figure}
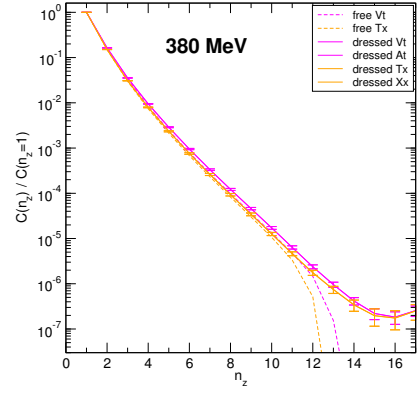

  \centering 
  \includegraphics[width=0.7\linewidth]{{{3}}} 
 \caption{Comparison of the full QCD correlators
  from the multiplets $E_3$ with with the correlators obtained
  on the same lattice with free noninteracting quarks. From Ref. \cite{R1}.}
\label{diff}
\end{figure}

Now we will compare the conserved charge density correlators in QCD 
at $T=380$ MeV with the
correlators calculated with free noninteracting quarks in Fig. \ref{diff}.
We observe that these correlators essentially coincide.\footnote{The diffractive
behavior of the free quark correlator in Fig. \ref{diff} at $n_z > 12$ is unphysical lattice artifact related to the finite lattice volume with  periodic boundary conditions in spatial directions \cite{GL}. It can be removed by combining periodic and antiperiodic boundary conditions, see Ref. \cite{Chiu}. Then
the full QCD correlator will coincide with the free quark loop correlator at all
$n_z$.} We conclude, that contrary to the mesonic correlators, the conserved charge density, propagating in $z$-direction, is described by free quarks in the
confining regime. The conserved charges do not see the confining force. This
fact explains why some cumulants of the fluctuations of conserved charges 
behave as if the quarks were free.The conserved quark density propagators
can be adequately represented by a free quark loop of Fig. \ref{loo}. 
This effect begins at rather low temperature, soon above the chiral restoration
temperature, see Fig. 6 of Ref. \cite{Chiu}, where the full QCD correlators from the multiplet $E_3$ are compared with the free quark gas correlators at $T= 193$ MeV.
Already at such temperature the difference between the full QCD correlators
$V_t$ and $A_t$
and the corresponding free quark gas correlators is  insignificant. This fact explains the previous observations that some cumulants of conserved charges approach the free quark gas value already at $T= 220$ MeV.

\section{Analogy to the quark-hadron duality observed at 
 $T=0$ in $e^+e^- \rightarrow hadrons$}

As the summary of the previous discussions we emphasize the following
facts. The mesonic propagators, both  spatial, discussed in the previous section, and temporal \cite{R3}, demonstrate that above $T_{ch}$ the propagating degrees of freedom are the color-singlet meson-like systems, where the chirally symmetric quarks
are connected into color-singlets by a confining electric string. These propagators are qualitatively different from propagators obtained with free
non-interacting quarks. This observation points out the confining regime, where
there are no deconfined gluons and the propagating physical degrees of freedom
are only the color singlets. This is consistent with other lattice observations,
discussed in the introduction. At the same time the conserved charges fluctuate
very soon above $T_{ch}$ as if the quarks were free, which means that the
conserved charges do not see the confining force. How could it be that
in the confining regime the conserved charges are not affected by confinement?

This situation is actually not new but is well familiar and studied in detail for many years.  Consider, as example, the  $e^+e^- \rightarrow hadrons$ annihilation
at $T=0$. The vacuum is confining, and consequently no
isolated quarks and gluons can be observed. Only color singlet hadrons
are registered in detectors. At the same time the inclusive cross-section
above the $\rho'$ bump, i.e., at the invariant mass of 2 GeV and above, is accurately reproduced by a free quark loop (plus some small perturbative corrections).  This fact does not imply that the quarks and gluons are deconfined.
It only means that the present observable is not sensitive to the confining 
infrared contributions \cite{Poggio:1975af}. This is called the quark-hadron duality.
For a detailed review of the issue see Ref. \cite{S}. 
At the same time the exclusive reactions, e.g. the hadron
formfactors, cannot be described by perturbation theory up to very large
momenta transfer, because they are always sensitive to the confining infrared 
contributions.

Then the analogy is rather direct. The mesonic color-singlet propagators above $T_{ch}$ but below the deconfinement temperature $T_d$ are directly sensitive to the confining force.
At the same time fluctuations of conserved charges are not sensitive to the confining force and behave as if the quarks were free. A possible microscopic picture
of the stringy fluid matter that is consistent with these properties is
outlined in the next section.

\section{Microscopic picture of the stringy fluid}

The first question that arises is how chiral symmetry could be restored in the
confining regime? This question was answered within the microscopic calculation in
the framework of the large $N_c$ manifestly confining and chirally symmetric model in $3+1$ dimensions \cite{LeYaouanc:1983huv,LeYaouanc:1984ntu,Adler:1984ri,Kocic:1985uq,Bicudo:1989sh,Bicudo:1989si,Llanes-Estrada:1999nat,Wagenbrunn:2007ie} that is similar to large $N_c$ QCD in
1+1 dimensions ('t Hooft model) \cite{tHooft:1974pnl}. It was demonstrated in Refs. \cite{gnw1,GNW}
that the chiral symmetry restoration in the confining regime happens because of Pauli blocking of the quark levels, required for the existence of the
chiral condensate, by the thermal excitations of quarks and antiquarks.
It was also obtained that the same Pauli blocking leads to the swelling of mesons
above $T_{ch}$ several times as compared their size in vacuum (in the chiral limit the "mesons" become infinitely large). This means that the matter is
a densely packed highly collective system of the overlapping color singlet "mesons" with a very small mean free path of the color singlet constituents.

While this model is only  some approximation to  real QCD, this qualitative microscopic picture suggests a coherent view of the stringy fluid matter that is consistent with all existing lattice data above $T_{ch}$ but below the deconfinement very smooth transition likely  centered around $T_d \sim 300$ MeV:

$\bullet$ It is a densely packed system of the overlapping color-singlet
objects.
 
$\bullet$ There are no deconfined gluons.

$\bullet$ It is a highly collective medium with a very small mean-free
path of the color-singlet constituents.

$\bullet$ The propagating in time degrees of freedom are only color-singlets,
in which the massless quarks and antiquarks are connected by the chromoelectric string.

Then, how could the fluctuations of conserved charges, where quarks behave
as if they were free, be accomodated into this picture?

It is actually  easy. Since the color singlet clusters strongly overlap,
Pauli principle requires the quark interchanges between the color singlet clusters, that should be crucially important. These quark interchanges
make the quarks quasi-free.  Notice that it is not a free quark propagation with time. This picture  should be considered as the Minkowski dual realization
of  the free quark loop z-propagation of conserved charge density  in Euclidean space.

\section{Conclusions}

In this paper we analysed the existing lattice data
and argued that the awailable information  above $T_{ch}$ from the 
correlators, equation of state, fluctuations of conserved charges, etc.
is consistent with the existence of the confining regime with restored chiral symmetry at temperatures $T_{ch} < T < T_d$. In particular, the fluctuations
of conserved charges that look like the matter was a gas of free quarks,
is because this observable is not sensitive to the infrared
confining contributions, in contrast to the mesonic correlators.
This is consistent with the well-known quark-hadron duality.
The stringy fluid matter can be viewed as a system of the overlapping
color-singlet clusters, with no deconfined gluons, and where the quark
interchanges between the clusters, required by Pauli principle, make the
quarks quasi-free. However, it is not a free propagation with time of the colored quarks.
Only the color singlets, where the quarks are connected by the confining
electric string, can propagate in time.
 
 The suggested microscopic picture of the
stringy fluid should be verified (falsified) in lat-
tice simulations. A very important aspect of this
picture is a huge swelling, by several times, of the
pion-like excitations above the chiral crossover, as
compared to the hadron gas. The size of these
objects can be studied by standard lattice sim-
ulations. Another crucial aspect of the stringy
fluid is the absence of deconfined gluons. This
issue can be addressed on the lattice via simula-
tions of the thermodynamical quantities (pressure,
energy density and entropy density) at Nc = 5, 7
in precisely the same manner as it was done for
Nc = 3. The appearance of the deconfined gluons
above Td must be accompanied by the $Nc^2$ scaling,
while in the stringy fluid the $Nc^1$ scaling should be
expected.

\section*{Acknowledgments}
The author is thankful to Larry McLerran and Owe Philipsen for discussions.
The  support through the grant PAT3259224 of the Austrian Science Fund (FWF) is acknowledged.

\end{document}